\documentclass[aip,reprint]{revtex4-1}

\usepackage{graphicx}
\usepackage{dcolumn}
\usepackage{bm}
\usepackage{braket}
\usepackage{subfigure}
\usepackage{graphicx}

\usepackage{color}

\draft 

\begin{document}


\title{A three-dimensional multimode lumped-element resonator for collective spin manipulation and dispersive readout}



\author{Zhuo Chen} \email{cz21@mail.ustc.edu.cn}
\affiliation{Hefei National Research Center for Physical Sciences at the Microscale and School of Physical Sciences, University of Science and Technology of China, Hefei 230026, China}
\affiliation{Shanghai Research Center for Quantum Science and CAS Center for Excellence in Quantum Information and Quantum Physics, University of Science and Technology of China, Shanghai 201315, China}
\author{Wenhua Qin}
\affiliation{Hefei National Research Center for Physical Sciences at the Microscale and School of Physical Sciences, University of Science and Technology of China, Hefei 230026, China}
\affiliation{Shanghai Research Center for Quantum Science and CAS Center for Excellence in Quantum Information and Quantum Physics, University of Science and Technology of China, Shanghai 201315, China}
\author{Hanyu Ren}
\affiliation{Hefei National Research Center for Physical Sciences at the Microscale and School of Physical Sciences, University of Science and Technology of China, Hefei 230026, China}
\affiliation{Shanghai Research Center for Quantum Science and CAS Center for Excellence in Quantum Information and Quantum Physics, University of Science and Technology of China, Shanghai 201315, China}
\author{Ziyi Liu}
\affiliation{Hefei National Research Center for Physical Sciences at the Microscale and School of Physical Sciences, University of Science and Technology of China, Hefei 230026, China}
\affiliation{Shanghai Research Center for Quantum Science and CAS Center for Excellence in Quantum Information and Quantum Physics, University of Science and Technology of China, Shanghai 201315, China}
\author{Kae Nemoto}
\affiliation{Okinawa Institute of Science and Technology Graduate University, Onna-son, Okinawa 904-0495, Japan}
\author{William John Munro}
\affiliation{Okinawa Institute of Science and Technology Graduate University, Onna-son, Okinawa 904-0495, Japan}
\author{Yingqiu Mao}
\affiliation{Hefei National Research Center for Physical Sciences at the Microscale and School of Physical Sciences, University of Science and Technology of China, Hefei 230026, China}
\affiliation{Shanghai Research Center for Quantum Science and CAS Center for Excellence in Quantum Information and Quantum Physics, University of Science and Technology of China, Shanghai 201315, China}
\author{Johannes Majer}\email{johannes@majer.ch}
\affiliation{Hefei National Research Center for Physical Sciences at the Microscale and School of Physical Sciences, University of Science and Technology of China, Hefei 230026, China}
\affiliation{Shanghai Research Center for Quantum Science and CAS Center for Excellence in Quantum Information and Quantum Physics, University of Science and Technology of China, Shanghai 201315, China}


\date{\today}

\begin{abstract}
We report a three-dimensional lumped-element multimode microwave resonator that enables homogeneous collective manipulation and dispersive readout of a macroscopic spin ensemble. By exploiting geometric symmetry, two antisymmetric modes with strongly suppressed cross-talk are engineered to spatially overlap and couple to the same ensemble at distinct frequencies. Using negatively charged nitrogen-vacancy ($\text{NV}^{-}$) centers in diamond at 28 mK, we observe collective strong coupling with a coupling strength of 5.0 MHz and demonstrate non-destructive dispersive readout via a detuned mode. The compact design, tunable coupling, and high field homogeneity make this resonator a versatile device for hybrid spin–photon systems and multimode solid-state quantum technologies.
\end{abstract}

\pacs{42.50.Pq}

\maketitle 

In hybrid quantum systems, the coherent exchange of quantum information is typically mediated by microwave photons \cite{RN261}. Achieving high-efficiency transfer requires operation in the strong-coupling regime, where the interaction rate exceeds dissipation in both subsystems. For individual spins, the magnetic-dipole coupling to a microwave resonator is intrinsically weak; however, coupling to an ensemble of $N$ spins enhances the interaction strength by a factor of $\sqrt{N}$ \cite{RN261}.
This collective enhancement has enabled strong coupling in a variety of cavity architectures, including superconducting coplanar waveguide resonators\cite{RN246,RN206,RN272,RN241,RN273} and three-dimensional microwave cavities  \cite{RN130,RN128,RN253,RN266,RN265,RN267,RN268,RN269,RN270,RN271}.

Many solid-state spin systems, such as negatively charged nitrogen-vacancy ($\text{NV}^{-}$) centers in diamond and donor spins in silicon, possess optical transitions. Spin readout is therefore often performed using optically detected magnetic resonance, which relies on optical pumping and fluorescence collection. While powerful, this approach is inherently irreversible and destructive. An alternative is dispersive microwave readout, widely employed in circuit quantum electrodynamics \cite{RN221,RN222}, in which the spin polarization induces a frequency shift of a detuned resonator without direct energy exchange. Dispersive readout enables continuous, non-destructive measurements and is compatible with feedback and real-time control\cite{RN150}. Achieving this functionality in compact hardware places stringent requirements on cavity geometry and mode structure, as illustrated in Fig. 1.

Implementing collective control and dispersive readout at distinct frequencies requires at least two spatially overlapping cavity modes that couple homogeneously to the same spin ensemble. In two-dimensional resonators, multimode operation can be realized using higher harmonics \cite{RN260} or cascaded coplanar waveguides \cite{RN255}; however, these approaches generally suffer from poor magnetic-field homogeneity. Three-dimensional cavities offer greater flexibility through mode engineering in the additional dimension, allowing control over the spatial distribution of electromagnetic energy. In particular, three-dimensional lumped-element resonators, such as bowtie \cite{RN253} and loop-gap geometries \cite{RN128}, can provide highly homogeneous magnetic fields with small mode volumes, enabling uniform Rabi frequencies and robust collective dynamics, as schematically illustrated in Fig. 1(b).
\begin{figure*}[htb]
\centering
    \subfigure[]{
		\begin{minipage}[t]{0.55\textwidth}
			\centering
			\includegraphics[width=9cm]{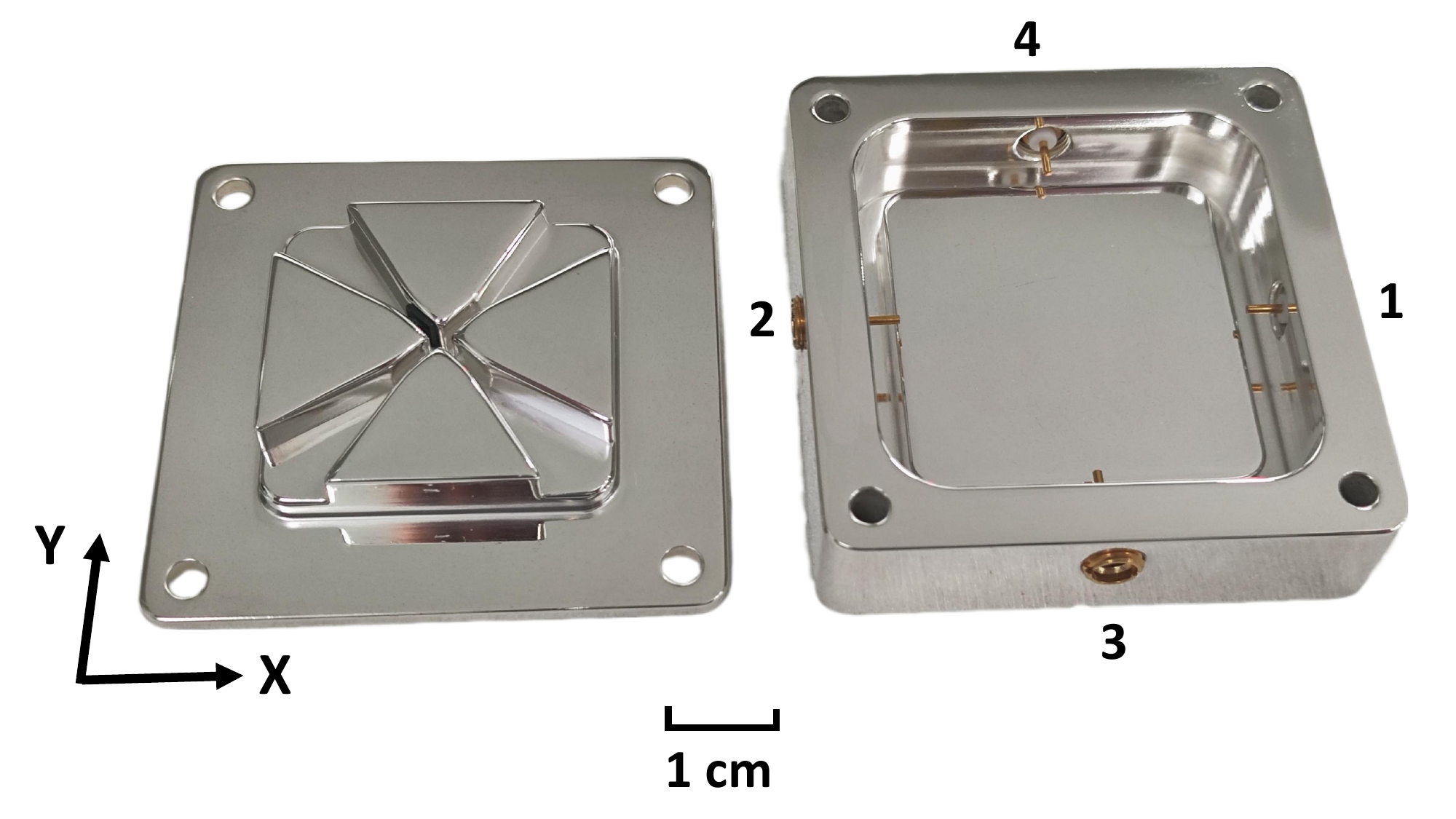}
            \label{3DXres_photo}
		\end{minipage}
    }
     \subfigure[]{
		\begin{minipage}[t]{0.4\textwidth}
			\centering
			\includegraphics[width=6.5cm]{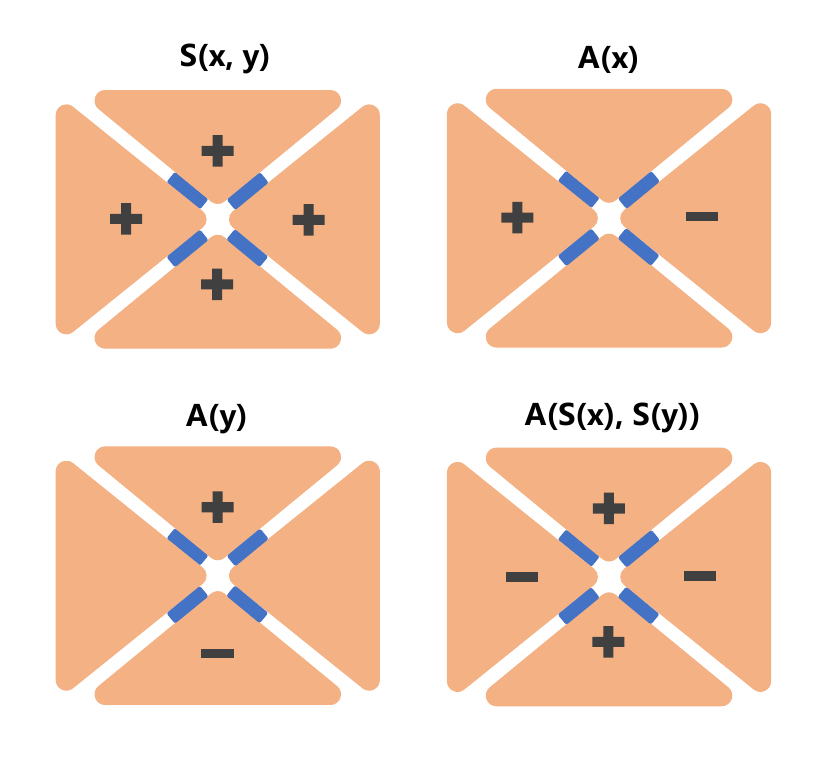}
            \label{modes_symmetry}
		\end{minipage}
    }
    \\
    \subfigure[]{
		\begin{minipage}[t]{0.55\textwidth}
			\centering
			\includegraphics[width=8cm]{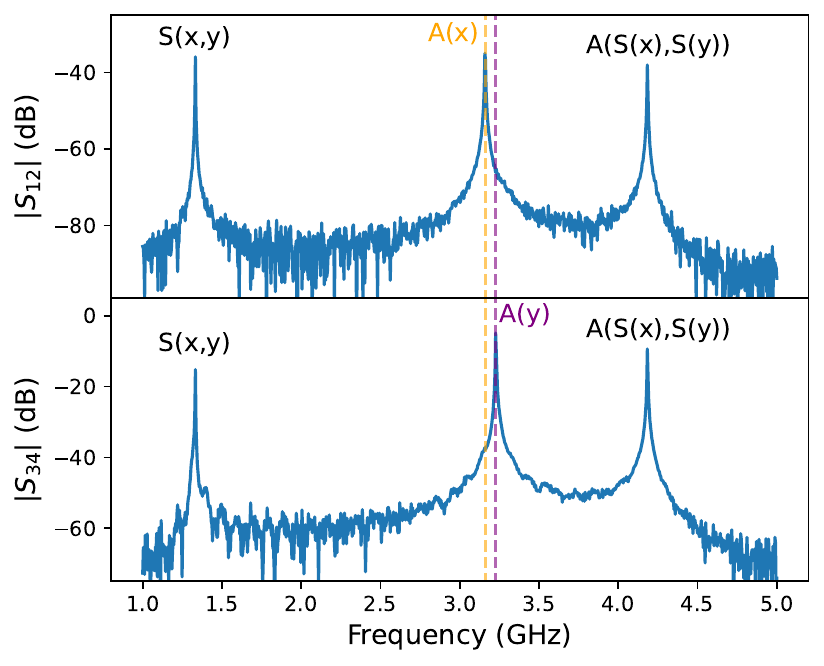}
            \label{s21xresyres}
		\end{minipage}
    }
    \subfigure[]{
		\begin{minipage}[t]{0.4\textwidth}
			\centering
			\includegraphics[width=6.5cm]{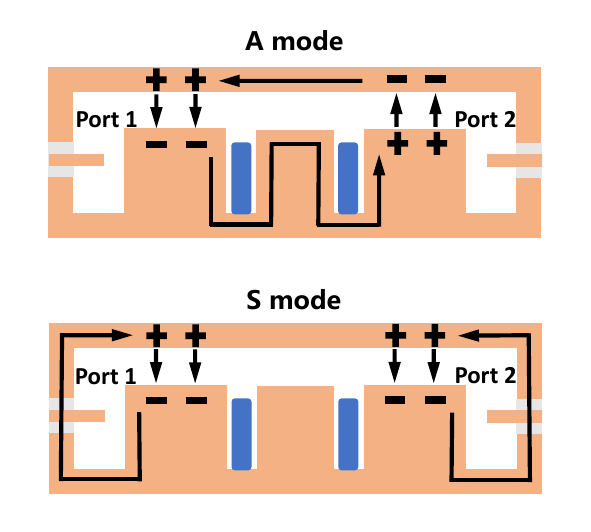}
            \label{current_flow}
		\end{minipage}
    }
    \caption{ (a) Photographs of the resonator base and enclosure.  Both parts are fabricated from oxygen-free copper and plated with silver; the surfaces are polished to mirror-level roughness, and  four microwave connectors are integrated into the enclosure. (b) Top-view schematics of the resonator modes classified by the symmetry of the electric potential.  The blue regions indicate locations where spin ensembles can be installed.  The plus and minus signs denote the electric potential. (c) Measured transmission spectra of the multimode resonator. The all-symmetric mode $S(x,y)$ produces a magnetic field distributed away from the center. The anti-symmetric modes $A(x)$ and $A(y)$ generate magnetic fields concentrated in the central slots, where the spin ensembles are located. For the hybrid mode $A(S(x),S(y))$, the magnetic field is distributed in both regions. The mode $A(y)$ has a higher resonance frequency because the wings along the $y$ axis are shorter than those along $x$. (d) Side views of the current flow for the symmetric and anti-symmetric modes. The modes are formed by a lumped capacitance between the wings and the top lid in series with a geometric inductance defined by the current path.}
\end{figure*}

Despite substantial progress in three-dimensional cavity designs, achieving fast collective manipulation together with non-destructive microwave readout of macroscopic spin ensembles remains challenging. This requires a combination of multimode operation, strong and homogeneous spin–photon coupling, and strong suppression of cross-talk between modes. Multimode cavity concepts with reduced cross-talk have been explored previously \cite{RN264,RN256}, but integrating these features with homogeneous collective coupling remains nontrivial. Here we address these challenges by introducing a three-dimensional lumped-element multimode resonator that combines a bowtie-inspired geometry with a symmetric cross-shaped layout, as shown in Fig. 1(a). The resulting structure supports two antisymmetric modes at distinct frequencies that spatially overlap and couple homogeneously to the same spin ensemble, while symmetry strongly suppresses intermode coupling, as evidenced by transmission spectroscopy in Fig. 1(c).

The multimode resonator consists of an inner structure formed by four triangular, wing-like elements that act as a large effective capacitance to the top lid while concentrating the microwave magnetic field into a small mode volume. The symmetric geometry enables the generation of a spatially homogeneous magnetic field at the resonator center. The device can be modeled at the circuit level as a lumped-element resonator with lumped capacitance $C$ and inductance $L$ yielding a resonance frequency $f = 1/\sqrt{LC}$, and the RF current runs near the metallic surface at a skin depth of several micrometers. The capacitance is primarily formed by the gaps between the wings and the top lid, and the geometric inductance arises from the current path around the X-shaped slot. The closed circuit of the capacitance in series with the inductance forms a lumped-element LC resonator. There are four microwave ports positioned near the wings. The spin ensembles are installed in the central slots, with up to four supported. 

There are four modes exhibited according to the electric potential symmetry. Considering the relative electric potential on the wings at resonance, starting from the lowest frequency, they are: the all-symmetric mode $S(x,y)$, in which all wings share the same potential; the $x$ anti-symmetric mode $A(x)$, where wings along the $x$ direction have opposite potentials while those along $y$ remain at zero; the $y$ anti-symmetric mode $A(y)$;  and the hybrid mode $A(S(x),S(y))$, respectively. 
Here, $S$ denotes ``symmetric'' and $A$ denotes ``anti-symmetric''. 
The side views of the current flow for the symmetric and anti-symmetric modes are shown in Fig.~\ref{current_flow}. 

Notably, aside from these four modes, other combinations of potential symmetry are forbidden.  This can be understood by considering the resulting potential differences in the structure.  For anti-symmetric modes resonant along the $x$ or $y$ direction, the center of the top enclosure acquires an electrical potential that differs from that of the bottom enclosure.  For the symmetric mode, the top and bottom enclosures share the same potential; these two configurations are therefore incompatible along orthogonal directions, leading to mode decoupling. 
In addition, the resonator does not support simultaneous anti-symmetric modes along both the $x$ and $y$ axes, because their current paths at the center are orthogonal. 


\begin{figure}
\centering
\includegraphics[scale=0.58]{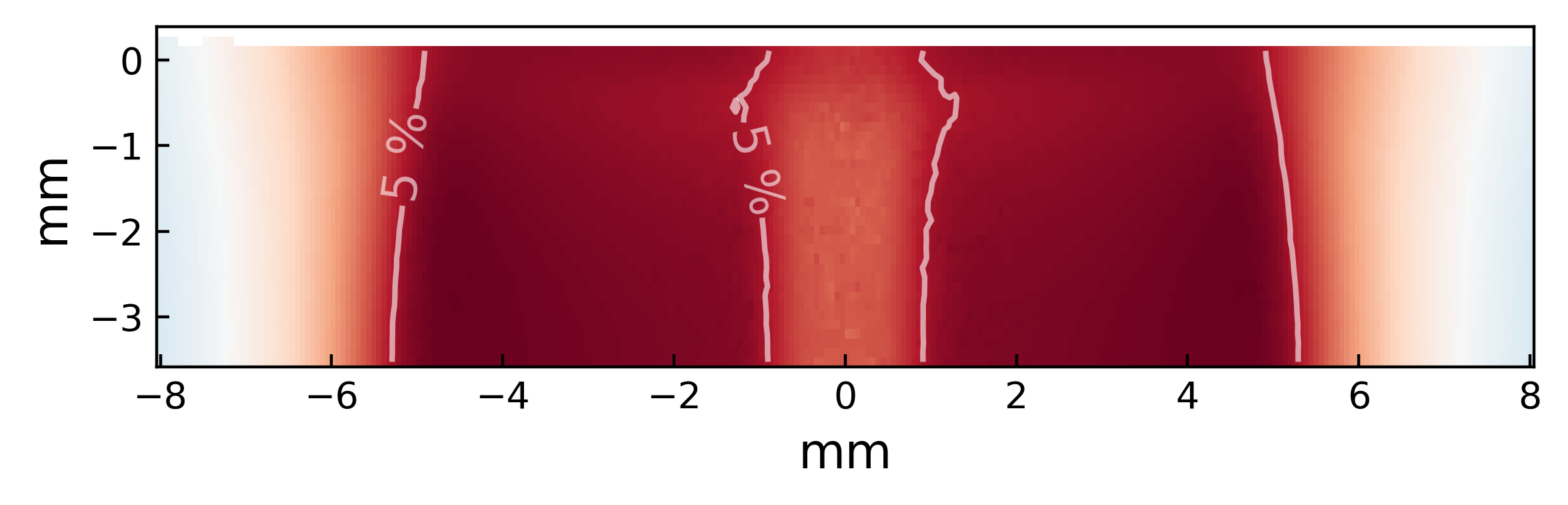}%
\\
\includegraphics[scale=0.58]{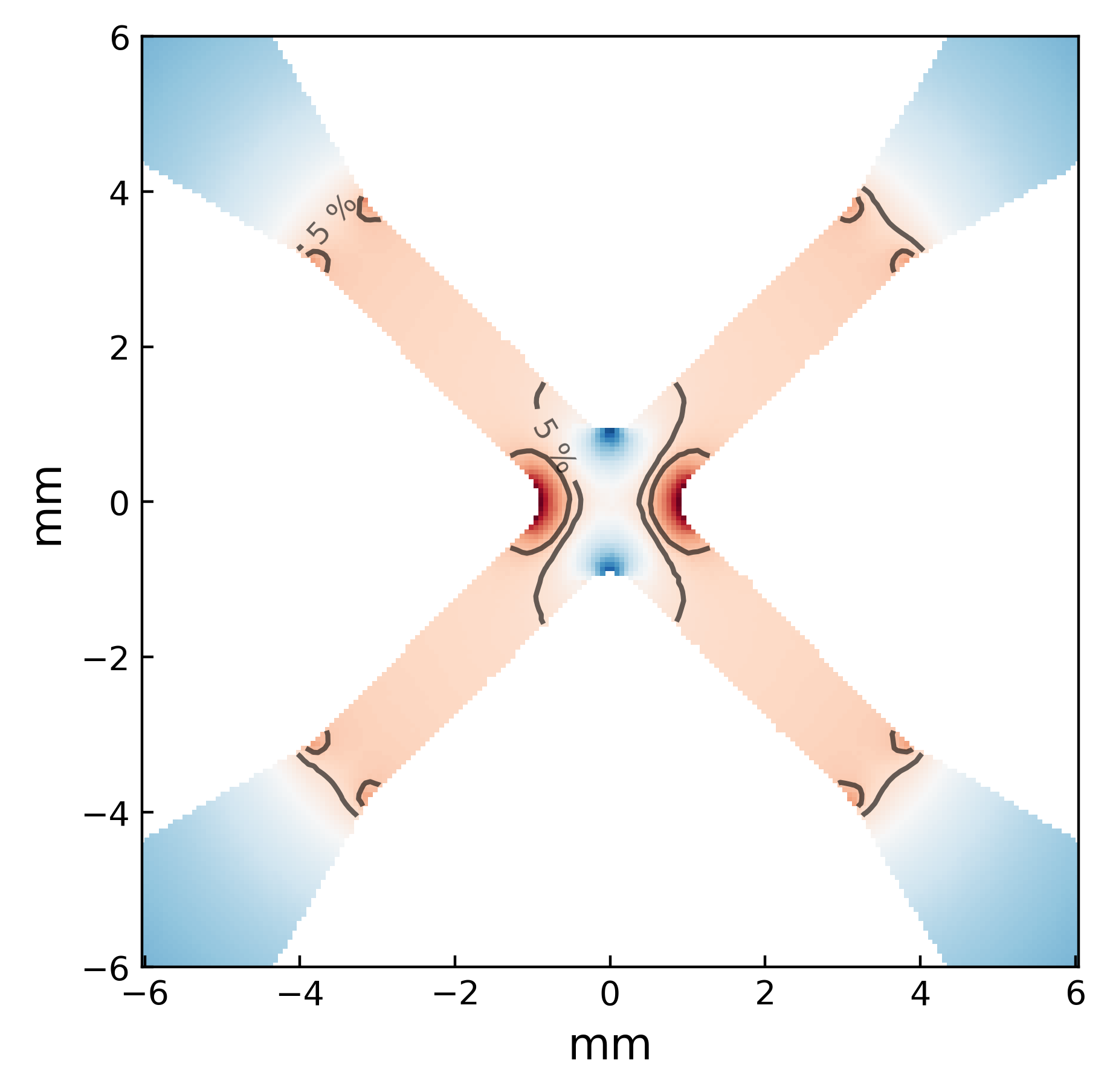}%
\caption{Simulated magnetic field distribution of the anti-symmetric mode $A(x)$. Top: side view, where the cross section is taken parallel to the $z$ axis and oriented at $45^\circ$ with respect to the $x$ axis. Bottom: top view, taken parallel to the $x$--$y$ plane. All contour lines correspond to a 5\% deviation in magnetic field strength relative to the field magnitude at the center of each slot.}
\end{figure}

When an anti-symmetric mode along the $x$ or $y$ axis is supported, both symmetric and anti-symmetric modes along the orthogonal direction are forbidden. The two anti-symmetric modes $A(x)$ and $A(y)$ are mutually decoupled and also decoupled from all other modes. In the transmission spectroscopy measurements (see Fig.~\ref{s21xresyres}), $S(x,y)$ and $A(S(x),S(y))$ appear in both $S_{12}$ and $S_{34}$, reflecting current loops distributed along both the $x$ and $y$ directions. In contrast, $A(x)$ appears only in $S_{12}$ and $A(y)$ appears only in $S_{34}$. The mode $A(y)$ has a higher resonance frequency because the wings along the $y$ axis are shorter than those along $x$, resulting in a smaller effective capacitance and thus a higher resonance frequency. To further suppress cross-mode leakage, the four microwave connectors are arranged orthogonally along the $x$ and $y$ axes. The geometric symmetry of the structure therefore provides strong suppression of cross-talk between different modes, which is essential for performing dispersive readout and strong coherent driving simultaneously while coupling spins to multiple resonances. During fabrication, polishing, and assembly, small geometric deviations are unavoidable. These imperfections partially break the symmetry and introduce residual cross-talk between modes. Experimentally, a cross-talk suppression exceeding 30~dB is nevertheless achieved.

The resonance frequency of the resonator is sensitive to geometric deviations introduced during fabrication and installation. These shifts can be compensated by in situ frequency tuning. By attaching dielectric films, such as Teflon, to the wing surfaces, the effective capacitance can be increased, thereby reducing the resonance frequency according to $f = 1/\sqrt{LC}$.  The coupling loss $\kappa_c$ can also be tuned by adjusting the depth of the connector center pin. The connectors are installed in the side wall with a screw head (see Fig.~\ref{3DXres_photo}); by turning the connector in or out, the coupling capacitance can be modified. In the undercoupled regime (small $\kappa_c$), the resonance quality factor $Q_{\mathrm{tot}} = Q_i + Q_c$ can reach values up to 1800. By inserting the connector further into the resonator, coupling rates $\kappa_c$ exceeding 6~MHz can be achieved, corresponding to a fast resonator with $Q_{\mathrm{tot}} < 300$. The coupling strengths at the four ports are individually tunable.

The surface roughness plays a critical role in determining the internal loss of the resonator due to the skin effect of the RF field. To reduce ohmic losses at the metallic surfaces, the resonator is plated with silver to a thickness of approximately 7~$\mu$m. Compared with oxygen-free copper, silver offers higher electrical conductivity and improved resistance to oxidation. Together, these design and fabrication features make the resonator a compact and versatile device for macroscopic microwave spin control and readout without optical elements, suitable for integration in hybrid quantum systems.

\begin{figure}
\includegraphics[scale=0.55]{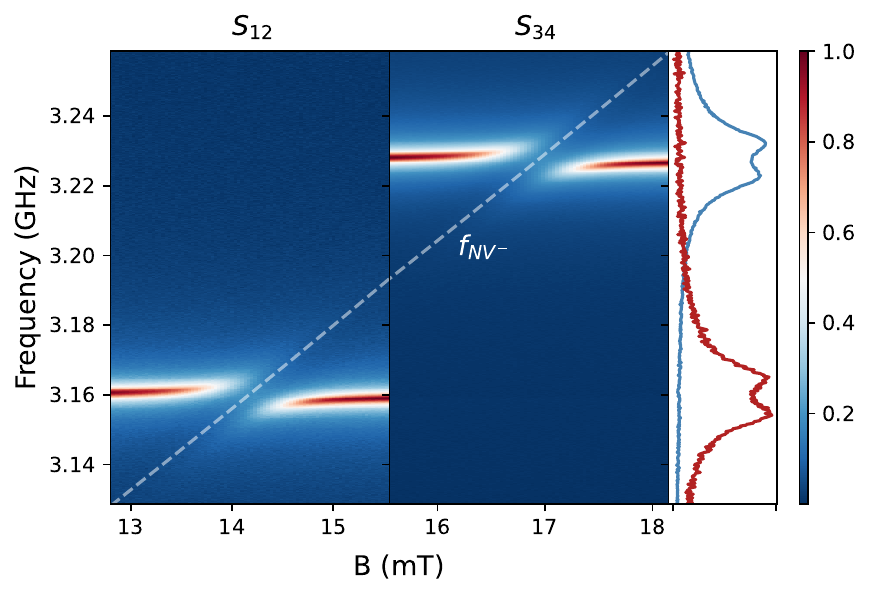}%
\caption{Collective strong coupling between multiple cavity modes and an $\text{NV}^-$ spin ensemble, involving the $m_s = 0 \rightarrow m_s = +1$ transition, whose frequency is tuned by an external static magnetic field. The left heatmap shows the measured transmission from port~1 to port~2 as a function of magnetic field, while the middle heatmap shows the transmission from port~3 to port~4. The right panel displays line cuts of the two transmission spectra when modes $A(x)$ and $A(y)$ are tuned into resonance with the spin ensemble at approximately 14.2~mT and 17.0~mT, respectively. The resonance frequency of mode $A(x)$ is 3.1598~GHz with a half width at half maximum (HWHM) of 0.995~MHz. The resonance frequency of mode $A(y)$ is 3.2275~GHz with an HWHM of 0.911~MHz. The white dotted lines indicate the calculated transition frequency of the $\text{NV}^-$ $m_s = +1$ state under a magnetic field applied along the $[100]$ crystallographic direction. All transmission amplitudes are normalized. The extracted collective coupling strength is 5~MHz for both modes $A(x)$ and $A(y)$.
\label{current_sweep}}%
\end{figure}

We insert a single diamond crystal into the resonator and place the device at the center of a superconducting three-dimensional Helmholtz vector magnet inside a dilution refrigerator. The diamond used in this experiment is a high-pressure high-temperature (HPHT) grown crystal with an $\text{NV}^-$ concentration of 55~ppm.  The diamond has a dark, nontransparent appearance and dimensions of $3.17 \times 3.15 \times 0.52$~mm$^3$. The estimated total number of $\text{NV}^-$ centers is $4 \times 10^{16}$, and all spins are homogeneously coupled to the resonator field. Using the vector magnet, the DC magnetic field is aligned along the $[100]$ crystallographic direction, such that the four NV sub-ensembles become degenerate. Neglecting hyperfine and spin--spin interactions, the Hamiltonian of the system is given by
\begin{eqnarray}
    H &=& \hbar \omega_{NV} \sum_{i}\sigma^{z}_{i} +\hbar  \omega_{c1} a^\dagger a + \hbar \omega_{c2} b^\dagger b + \nonumber\\
        &\,& \hbar \sum_{i} \left[ g_{1}\left( \sigma^{+}_{i} a + \sigma^{-}_{i} a^\dagger \right) 
        + g_{2} \left( \sigma^{+}_{i} b + \sigma^{-}_{i} b^\dagger \right)\right],
\end{eqnarray}
where $\omega_{c1}$ and $\omega_{c2}$ are the resonance frequencies of modes $A(x)$ and $A(y)$, respectively. The two modes couple strongly and homogeneously to the spin ensemble, enabling collective coherent driving of the spins with a uniform Rabi frequency. Now the frequency of mode $A(x)$ is $\omega_{c1}/2\pi = 3.1598$~GHz, with a half width at half maximum (HWHM) of $\kappa_{1}/2\pi = 0.995$~MHz while the frequency of mode $A(y)$ is $\omega_{c2}/2\pi = 3.2275$~GHz, with an HWHM of $\kappa_{2}/2\pi = 0.911$~MHz.  The inhomogeneous broadening of the spin ensemble is estimated to be 6~MHz (HWHM). Clear avoided crossings are observed in the transmission spectrum, as shown in Fig.~\ref{current_sweep}.

In a drive--readout scheme, the drive resonator is tuned into resonance with the spins to enable energy exchange, while the readout resonator is kept far detuned to minimize readout-induced spin relaxation and dephasing. The dispersive readout relies on the condition that the detuning between the spins and the readout resonator is much larger than the collective coupling strength.  In this regime, the readout resonator does not exchange photons directly with the spins. Instead, the presence of the spin ensemble imparts a spin-state-dependent dispersive shift to the resonant frequency. To first order, the relationship between the spin polarization $\langle S_z \rangle$ and the frequency shift $\chi$ is given by

\begin{equation}
    \chi = \frac{2g^2\braket{S_{z}}}{\omega_{c2}-\omega_{NV}}, 
\end{equation} 
where $g$ denotes the single-spin coupling strength. $\text{NV}^-$ centers are spin-1 defects, and both the $m_s = 0 \rightarrow m_s = -1$ and $m_s = 0 \rightarrow m_s = +1$ transitions couple to the resonator and contribute to the dispersive shift. In this experiment, the readout mode $A(y)$ is first characterized at zero external magnetic field. In this case, the $m_s = \pm 1$ states are degenerate at a transition frequency of 2.87~GHz, corresponding to the zero-field splitting. With all spins far detuned, mode $A(y)$ can be treated as a bare cavity resonance. 

\begin{figure}
\includegraphics[scale=0.52]{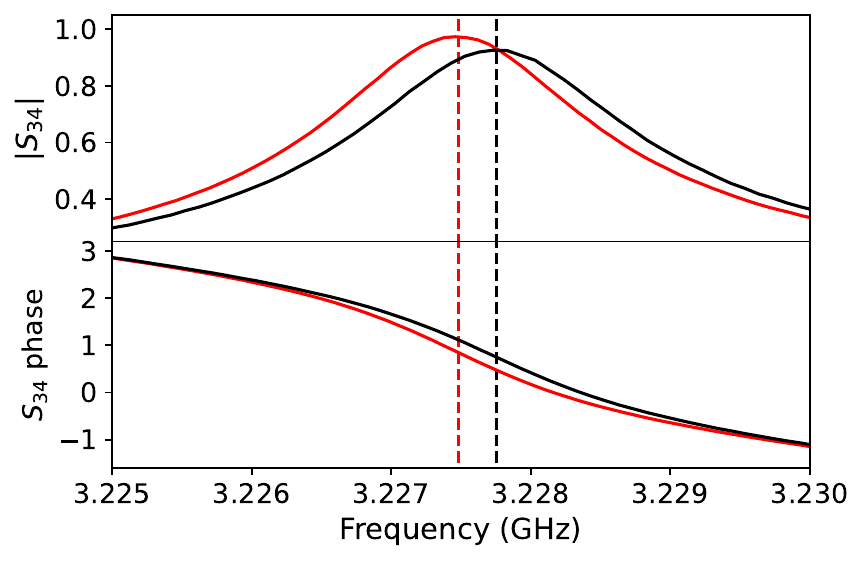}%
\caption{Measured transmission amplitude and phase of the readout mode $A(y)$ from port~4 to port~3 for different spin transition frequencies. Red traces correspond to zero external magnetic field, where the $\text{NV}^-$ spin transition frequency is 2.87~GHz. Black traces correspond to the case where the spins are tuned into resonance with mode $A(x)$ at 3.1598~GHz. The detuning between mode $A(y)$ and the spin ensemble is 67.7~MHz, which is much larger than the collective coupling strength. A clear dispersive frequency shift of $\chi = 0.27$~MHz is observed for mode $A(y)$.
\label{fit_coupling}}%
\end{figure}

Subsequently, an external static magnetic field is applied, and the Zeeman effect lifts the degeneracy of the $m_s = \pm 1$ states. The transition frequency of the $m_s = +1$ state increases and is tuned into resonance with the drive mode $A(x)$, such that $\omega_{c1}/2\pi = \omega_{NV}/2\pi = 3.1598$~GHz. Throughout this process, the transmission of the readout mode $A(y)$ is continuously monitored using a weak probe tone.  During all measurements, the spins remain well polarized in the ground state at a temperature of 28~mK.  From Fig.~\ref{fit_coupling}, a dispersive shift of $\chi = 0.27$~MHz is observed. The detuning between the spin ensemble and mode $A(y)$ is 67.7~MHz. Using Eq.~(2), the extracted collective coupling strength is 5.0~MHz. 

The number of $\text{NV}^-$ spins participating in the $m_s = +1$ state is estimated to be $2 \times 10^{16}$.  From the relation $g_{\mathrm{col}} = g_0 \sqrt{N}$, the corresponding single-spin coupling strength is $g_0/2\pi = 35$~mHz. As shown in Fig.~\ref{fit_coupling}, when the resonator is dispersively coupled to the spin ensemble, both the transmission amplitude and quality factor exhibit a slight reduction compared to the bare resonance, due to residual spin-induced loss. 
Because the inhomogeneous broadening of the spin ensemble is much smaller than the detuning between the spins and the readout mode, the observed dispersive shift is insensitive to the inhomogeneous broadening.

Next the dispersive readout exploits the intrinsic nonlinearity of the spin ensemble. The dispersive shift is directly proportional to the spin excitation, making it well suited for probing nonlinear dynamics in systems with large spin numbers and beyond the low-excitation regime. 
This capability enables the investigation of collective nonlinear spin phenomena, such as superradiance and superabsorption.

In conclusion, we design and experimentally demonstrate a three-dimensional multimode X-shaped lumped-element resonator that supports two spatially overlapping modes with homogeneous coupling to a macroscopic spin ensemble at its center. Operating near 3 GHz, the resonator exploits geometric symmetry to achieve strong suppression of intermode cross-talk, enabling collective strong coupling and non-destructive dispersive readout within a single compact device. Using an ensemble of $\text{NV}^{-}$ centers in diamond at 28~mK, we observe collective coupling strengths of 5.0~MHz and resolve a clear dispersive frequency shift of a detuned readout mode. These results demonstrate that the 3D multimode X-shaped resonator enables collective spin manipulation and microwave readout of macroscopic spin ensembles, with potential applications in hybrid spin–photon systems and solid-state quantum technologies.

\begin{acknowledgments}
We thank Zheheng Yuan, Tao Rong, Tao Jiang and Victor R. Garcia for help in developing hardware and software in early stage experiment setup, and also we appreciate Profs. Yizheng Zhen, Barry C. Sanders and Jörg Wrachtrup for inspiring discussions. This work has been supported by the Quantum Science and Technology-National Science and Technology Major Project (No. 2024ZD0301200), National Natural Science Foundation of China (No. W2431001). Y. M. acknowledges support from the Fundamental Research Funds for the Central Universities (No. WK9990000159).
\end{acknowledgments}

\bibliography{my-bib}

\end{document}